\def\draftversion{false}
\def\showall{true}  

\RequirePackage{ifthen}
\ifthenelse{\equal{\draftversion}{true}}{
  \documentclass[aps,prl,,10pt,galley,amsmath,amssymb,
                 longbibliography,superscriptaddress,nofootinbib]{revtex4-2}

}{
  \documentclass[aps,prl,10pt,twocolumn,amsmath,amssymb,floatfix,
    longbibliography,superscriptaddress,nofootinbib]{revtex4-2}
}

\usepackage{amsmath,amssymb,amsfonts,mathtools}
\usepackage{amsthm}
\usepackage{nicefrac}
\usepackage{graphicx}
\usepackage{hhline}
\usepackage[utf8]{inputenc}
\usepackage{subfigure}
\usepackage{xcolor}
\usepackage{bbm}
\usepackage{ulem}
\usepackage{mathrsfs}
\usepackage{cancel}
\usepackage{dsfont}

\usepackage{soul}  
\usepackage{hyperref}
\hypersetup{
    colorlinks=true,       
    linkcolor=blue,          
    citecolor=blue,       
    filecolor=magenta,      
    urlcolor=blue          
}

\ifthenelse{\equal{\draftversion}{true}}{
  \marginparwidth 2.7in
  \marginparsep 0.5in
  \newcounter{comm} 
  \def\commnext{\stepcounter{comm}}
  \def\commtext{{\bf\color{blue}[\arabic{comm}]}}
  \def\commmar{{\bf\color{blue}[\arabic{comm}]}}
  \def\stm#1{\commnext\marginpar{\small ST\commmar: #1}\commtext}
  \def\ism#1{\commnext\marginpar{\small IS\commmar: #1}\commtext}
   \def\xlm#1{\commnext\marginpar{\small XL\commmar: #1}\commtext}
  
  \def\parsedate #1:20#2#3#4#5#6#7#8\empty{20#2#3/#4#5/#6#7}
  \def\moddate{\expandafter\parsedate\pdffilemoddate{\jobname.tex}\empty}
  \newcommand{\mydate}{\date[Last modified ]{\moddate}}
  \def\tenxl#1{\textcolor{blue}{#1}}   
\def\tents#1{\textcolor{orange}{#1}}   
\def\tenti#1{\textcolor{red}{#1}}   
}{
  \def\stm#1{}
  \def\ism#1{}
  \def\xlm#1{}
  
  \newcommand{\mydate}{\date{\today}}
  \def\tenxl#1{{#1}}   
\def\tents#1{{#1}}   
\def\tenti#1{{#1}}   
}

\ifthenelse{\equal{\showall}{false}}{
  \renewcommand\includegraphics[2][]{{\bf Figure not shown.}}
  \renewcommand\input[1][]{{\bf Not shown\ }}
}{}

\newcommand{\beq}{\begin{equation}}
\newcommand{\eeq}{\end{equation}}
\newcommand{\bea}{\begin{eqnarray}}
\newcommand{\eea}{\end{eqnarray}}
\newcommand{\beas}[1]{\begin{subequations}\eql{#1}\begin{eqnarray}}
\newcommand{\eeas}{\end{eqnarray}\end{subequations}}

\usepackage{tikz}

\newcommand{\eq}[1]{Eq.~(\ref{eq:#1})}

\newcommand{\eqs}[2]{Eqs.~(\ref{eq:#1}) and (\ref{eq:#2})}

\newcommand{\Eqr}[2]{Equations~(\ref{eq:#1}-\ref{eq:#2})}

\newcommand{\fref}[1]{Fig.~\ref{fig:#1}}

\ifthenelse{\equal{\draftversion}{true}}{
  \newcommand{\eql}[1]{\label{eq:#1}\hbox{\Red{\small\;\;[#1]}}}
  \newcommand{\figl}[1]{\label{fig:#1}\Red{\small\;\;[Fig:~#1]}}
  \newcommand{\secl}[1]{\label{sec:#1}\Red{\small\;\;[Sec:~#1]}}
  \newcommand{\tabl}[1]{\label{tab:#1}\Red{\small\;\;[Tab:~#1]}}
}{}
\ifthenelse{\equal{\draftversion}{false}}{
  \newcommand{\eql}[1]{\label{eq:#1}}
  \newcommand{\figl}[1]{\label{fig:#1}}
  \newcommand{\secl}[1]{\label{sec:#1}}
  \newcommand{\tabl}[1]{\label{tab:#1}}
}{}

\def\ek{\tilde \epsilon_\kk}
\def\rdot{\mathbf{\dot{r}}}
\def\kdot{\mathbf{\dot{k}}}

\def\OO{\mathbf{\Omega_\kk}}

\newcommand{\Real}[1]{\rm Re\,}
\newcommand{\Imag}[1]{\rm Im\,}
\newcommand{\tr}[1]{\rm tr\,}
\newcommand{\Tr}[1]{\rm Tr\,}

\def\Tr{\mathrm{Tr}}

\def\k{{\bf k}}
\def\m{{\bf m}}
\def\rr{{\bf r}} 

\def\0{{\bf 0}}
\def\E{\boldsymbol{E}}
\def\B{\boldsymbol{B}}
\def\I{\boldsymbol{I}}
\def\j{\boldsymbol{j}}

\def\kk{{\bf k}}
\def\rr{{\bf r}}

\def\dk{[d\kk]}

\def\dk{[d\k]}

\def\la{\langle\kern-2.5pt\langle}
\def\ra{\rangle\kern-2.5pt\rangle}
\def\vt{\vert\kern-1.5pt\vert}
\begin{document}

\title{
  Electrical magnetochiral anisotropy
  in trigonal tellurium from first principles}

\author{Xiaoxiong Liu} \affiliation{Physik-Institut, Universit\"at
  Z\"urich, Winterthurerstrasse 190, CH-8057 Z\"urich, Switzerland}
\affiliation{Shenzhen Institute for Quantum Science and Engineering and Department of Physics, Southern University of Science and Technology (SUSTech), Shenzhen 518055, China}
\affiliation{Quantum Science Center of Guangdong-Hong Kong-Macao Greater Bay Area (Guangdong), Shenzhen 518045, China}
\affiliation{Shenzhen Key Laboratory of Quantum Science and Engineering, Shenzhen 518055, China}
\affiliation{International Quantum Academy, Shenzhen 518048, China}

\author{Ivo Souza} \affiliation{Centro de F{\'i}sica de Materiales,
  Universidad del Pa{\'i}s Vasco, 20018 San Sebasti{\'a}n,
  Spain} \affiliation{Ikerbasque Foundation, 48013 Bilbao, Spain}

\author{Stepan S. Tsirkin} \affiliation{Centro de F{\'i}sica de Materiales,
  Universidad del Pa{\'i}s Vasco, 20018 San Sebasti{\'a}n,
  Spain} \affiliation{Ikerbasque Foundation, 48013 Bilbao, Spain}

  \begin{abstract}
  Structural chirality gives rise to characteristic responses that
  change sign with the handedness of the crystal structure. One
  example is electrical magnetochiral anisotropy (eMChA), a change in
  resistivity that depends linearly on the applied current and on the
  magnetic field.  Motivated by recent measurements of a strong eMChA
  in $p$-doped trigonal tellurium, we carry out an \textit{ab initio}
  study of the eMChA response in this material as a function of
  temperature and doping concentration.  We use the semiclassical
  Boltzmann transport formalism within the constant relaxation-time
  approximation to express the bulk eMChA response tensor in terms of
  the energy dispersion, intrinsic magnetic moment, and Berry
  curvature of the conduction Bloch states.  We find that the orbital
  Zeeman coupling dominates the calculated response, with smaller
  contributions coming from the spin Zeeman coupling and from the
  Berry curvature, and that the effect is maximal when both the
  current and magnetic field are along the trigonal axis. 
  The calculated data shows a reasonable agreement with the
   experiments. We provide the open-source code
   to facilitate further \textit{ab initio} studies of eMChA in other
   materials.
%
%
\end{abstract}
\mydate
\maketitle

\ifthenelse{\equal{\draftversion}{true}}{Color code: \tents{Stepan}, \tenti{Ivo}, \tenxl{Xiaoxiong}}{}

\secl{intro}

\textit{Introduction.}
The electrical transport properties of nonmagnetic crystals change
under an applied magnetic field, leading to different phenomena that
can be generally described as an expansion of the current in powers of
the electric and magnetic fields $\E$ and $\B$,
\beq
j_a = \sigma_{ab}(\B) E_b + \sigma_{abc} (\B) E_b E_c + \ldots\,.
\eql{j-E-expansion}
\eeq
The ordinary and planar Hall effects~\cite{hurd-book72}, as well as
the ordinary magnetoresistance~\cite{pippard1989magnetoresistance},
are all contained in the first term.  Electrical responses that are
quadratic in $\E$ (second term) become allowed in
noncentrosymmetric media, and have recently attracted a great deal of
attention. One example is the nonlinear anomalous Hall effect, a
transverse response that occurs in the absence of an external magnetic
field and without requiring magnetic
  order~\cite{deyo-arxiv09,PhysRevLett.115.216806,kang2019nonlinear}.

Another example of an $\E^2$ response is
electrical magnetochiral anisotropy
(eMChA)~\cite{Rikken01,atzori-chir21},
an anomalous magnetoresistance that is bilinear in $\B$ and in
the applied current~$\I$.
The resistance of a material exhibiting eMChA takes the form
\beq
R(B,I) = (1 + \beta B^2 + \gamma I B)R_0\,,
\eql{R-B-I}
\eeq
where the coefficients $\beta$ and $\gamma$ describe ordinary
  magnetoresistance and eMChA, respectively (the tensorial character
  of the various quantities is ignored for now).  The coefficient
$\gamma$ vanishes in centrosymmetric systems, and in chiral media it
is equal in magnitude and opposite in sign for right- and left-handed
enantiomers. As eMChA is not restricted to chiral systems, it is
sometimes referred to by other names such as bulk
current-rectification effect~\cite{ideue-natphys17}, nonreciprocal
resistance~\cite{ideue-natphys17,yoshimi-prb22}, and unidirectional
magnetoresistance~\cite{tokura-natcomms18,calavalle2022gate}.

The eMChA has been measured in several
chiral systems including bismuth helices~\cite{Rikken01},
single-walled carbon nanotubes~\cite{krstic-jcp02},
molecular conductors~\cite{pop2014electrical},
noncentrosymmetric transition metal dichalcogenide~\cite{emchaWTe2}, and trigonal
tellurium~\cite{Rikken19,Hirobe22,calavalle2022gate,niu2023tunable},
as well as in the nonchiral polar semiconductor
BiTeBr~\cite{ideue-natphys17}. 
%
Similar nonreciprocal responses
have also been reported both in Josephson junctions
\cite{JJ-SDE-2022,JJ-SDE-2022-NiTe2} and in junction-free
superconductors \cite{MoS2-SDE,NbTaV-SDE,SDE-2022-NbSe2}.

Recently, Calavalle \textit{et al.}~\cite{calavalle2022gate} reported
the manipulation and detection of eMChA
in tellurium nanowires at low
temperatures.
Motivated by this work and by that of Refs.~\cite{Rikken19,Hirobe22}, here we
undertake an \textit{ab initio} study of the eMChA response in
$p$-doped trigonal tellurium ($p$-Te). As it
crystallizes in the simplest chiral structure with only three atoms
per cell, elemental tellurium is an ideal prototype material for
performing 
bulk \textit{ab initio} calculations of eMChA.

\textit{Bulk tensorial description.} To formulate eMChA as a bulk
response, we divide \eq{R-B-I} by the cross-section area $A$ of the
sample, converting from resistance $R$ and current $I$ to resistivity
$\rho$ and current density $j$. Dropping the
ordinary magnetoresistance term and defining $\gamma'=\gamma A$, we obtain
\beq
\rho(B,j)=(1+\gamma' j B)\rho_0\,,
\eql{rho-B-j}
\eeq
where the tensorial character is again ignored.

While experimentally one applies currents and measures voltages,
microscopic theory deals instead with the current density induced by
the applied fields, as in \eq{j-E-expansion}.  We expand $\sigma(\B)$
in powers of $\B$, leaving only terms that are proportional to $E$ and
$E^2B$ \ to find
\beq
j_a =
\sigma_{ab}E_b +
\sigma_{abcd}E_b E_c B_d\,,
\eql{j-E-EEB}
\eeq
where $\sigma_{ab}$ is the linear Ohmic conductivity, $\sigma_{abcd}$
is the eMChA magnetoconductivity, and a summation over repeated
Cartesian indices is implied.  Solving for $\E$ as a function of $\B$
and $\j$ to order $\B\j^2$, one arrives at the following tensorial
generalization of \eq{rho-B-j},
\beq
\rho_{ae}(\B,\j)=(\delta_{ab}+\gamma'_{abcd} j_c B_d)\rho_{be}\,,
\eql{rho-Bj}
\eeq
where $\rho_{ac}\sigma_{cb}=\delta_{ab}$, and
\beq
\gamma'_{abcd}=-\rho_{aa'}\sigma_{a'bb'd}\rho_{b'c}
\eql{gamma-prime}
\eeq
is the bulk eMChA tensor. Thus, the main task of a predictive bulk
theory of eMChA is to evaluate the conductivity tensors $\sigma_{ab}$
and $\sigma_{abcd}$ defined by \eq{j-E-EEB} that are needed to
evaluate $\gamma'_{abcd}$ from \eq{gamma-prime}. 
Note that while
$\sigma_{abcd}$ has the intrinsic permutation symmetry
$\sigma_{abcd}=\sigma_{acbd}$, the relation
$\gamma'_{abcd}=\gamma'_{acbd}$ only holds when the Cartesian axes are
chosen as principal axes of the Ohmic resistivity $\rho_{ab}$.


\secl{general}

\textit{Boltzmann-transport theory of eMChA.}
To develop a microscopic theory of eMChA in nonmagnetic conductors, we
resort to the
semiclassical Boltzmann formalism with Berry-curvature and
  spin/orbital-moment corrections, which are widely used to describe linear
and nonlinear (magneto)transport in
solids~\cite{RevModPhys.82.1959,gao-fp19}.  Within the constant
  relaxation time approximation adopted below, this formalism accounts
  for intrinsic band structure effects but leaves out extrinsic
  effects such as skew scattering and side jump.

In the Boltzmann formalism, the current density takes the form
\beq
{\bf j} = - \int \dk \, D \rdot f\,,
\eql{BB-j}
\eeq
where the integral is over the Brillouin zone,
$\dk\equiv d^3k/(2\pi)^3$, $\rdot$ is the electron velocity, $f$ is
the nonequilibrium distribution function, and $D\equiv 1+\OO  \cdot\B$,
with $\OO$ the Bloch-state Berry curvature. We use units where
$e=\hbar=m_e=1$, and omit band summations.

We determine $f$ by solving the Boltzmann transport equation in the
constant relaxation-time approximation,
\beq
\kdot\cdot \nabla_{\kk} f = \frac{f_0(\ek)-f}{\tau}\,,
\eql{boltzmann}
\eeq
where $f_0(\ek)$ is the equilibrium Fermi-Dirac distribution function
evaluated at the Zeeman-shifted band energy
$\ek=\epsilon_\kk -\bf{m}_\kk \cdot \B$, with $\m_\kk$ the total (spin
plus orbital) magnetic moment $\m_\kk$ of a Bloch
state~\cite{gao-prb17}.
The velocities $\rdot$ and
$\kdot$ in real space and in reciprocal space
satisfy the coupled equations~\cite{RevModPhys.82.1959}
\begin{subequations}
\bea
\rdot &= \nabla_\kk \ek -\kdot \times \OO \eql{BB-r}\,, \\
\kdot &= -\E - \rdot \times \B  \eql{BB-k}\,,
\eea
\label{BB-rk}
\end{subequations}
which can be decoupled as
\begin{subequations}
\bea
D\rdot &= \nabla_\kk \ek +\E \times \OO+
\left( \nabla_\kk \ek \cdot\OO\right)\B  \eql{BB-Dr}\,, \\
-D\kdot &=  \E+ \nabla_\kk \ek \times\B + \left(\E\cdot\B\right)\OO~.
\eql{BB-Dk}
\eea
\label{BB-Drk}
\end{subequations}
%
Next, we expand $f$ in \eq{boltzmann} as
$f = \sum_{n=0}^\infty f_n \tau^n$ to obtain the recursive relation
\beq
f_{n+1} = -\tau\kdot\cdot\nabla_\kk f_n~,
\eql{tau-f}
\eeq
which is solved starting from $f_0=f_0(\ek)$ using \eq{BB-Dk} for
$\dot\kk$. Finally, the current density is obtained as a power series
in $\E$ and $\B$ by inserting the resulting expression for $f$ into
\eq{BB-j}, together with \eq{BB-Dr} for~$\dot\rr$.

At linear order in $\E$, the above procedure yields the familiar
result for the Ohmic conductivity,
\beq
\sigma_{ab} = \tau\int\dk \, v_a v_b f_0'\,,
\eql{sigma-ab}
\eeq
where
$f_0^\prime = d
f_0(\epsilon)/d\epsilon\vert_{\epsilon=\epsilon_\kk}$. The derivation
of the expression at order $\E^2\B$ for the eMChA conductivity
$\sigma_{abcd}$ is outlined in the Supplementary Material~\cite{SM}.
Here, it suffices to note that in nonmagnetic conductors the only
nonvanishing contribution comes from $f_2$, and so it is proportional
to $\tau^2$. In particular, there is no contribution coming from the
$\B$ dependence of the chemical potential, which is an even function
of $\B$ in crystals with time-reversal symmetry~\cite{gao-prb17}. The
resulting expression
\beq
\sigma_{abcd}=
\sigma^{\rm Z}_{abcd} + \sigma^{\rm \Omega}_{abcd}
\eql{sigma-MC}
\eeq
comprises the Zeeman (Z) and Berry curvature $(\Omega)$ terms
\beq
\sigma^{\rm Z}_{abcd}= \tau^2 \int \dk \,
\left( v_{abc}m_d - v_c \partial_{ab} m_d   \right)f_0^\prime
\eql{sigma-Z}
\eeq
and
\begin{align}
\sigma^{\rm \Omega}_{abcd} = -\tau^2 &\int \dk \,
\Big[ 2 v_{ab} v_c \Omega_d + v_a v_c\partial_b\Omega_d \nonumber\\
&-\delta_{cd} (2v_{ae} v_b \Omega_e  - v_a v_e\partial_b\Omega_e )\nonumber\\
&-\delta_{ad}(v_{be} v_c\Omega_e +  v_c v_e\partial_b\Omega_e)\nonumber\\
&-\varepsilon_{edf}\varepsilon_{abg} v_f \partial_e \Omega_g v_c  
\Big] f_0^\prime
\eql{sigma-mc'}
\,,
\end{align}
where $\partial_a\equiv\partial/\partial_{k_a}$,
$v_a \equiv \partial_a \epsilon_\kk$ is the band velocity,
$v_{ab} \equiv \partial_{ab} \epsilon_\kk$ is the inverse
effective-mass tensor, and
$v_{abc} \equiv \partial_{abc} \epsilon_\kk$. 
Note that only the part of $\sigma_{abcd}$ that is invariant under permutations 
of  indices $b$ and $c$ contributes to the physical effect, while 
in \eqs{sigma-Z}{sigma-mc'} the symmetry under $b\leftrightarrow c$ is not explicitly seen.
Therefore, we further use the combinations $(\sigma_{abcd}+\sigma_{acbd})/2$
for the 
component $\sigma_{abcd}$.
The Zeeman response can
be further separated into spin and orbital parts, yielding three types
of contributions in total. The spin-Zeeman response requires spin-orbit coupling, while the
orbital-Zeeman and Berry curvature responses are present even without
spin-orbit coupling.

In the above derivation, we did not consider the corrections to the
orbital moment and Berry curvature due to applied
fields~\cite{gao-prl14,gao-fp19}.
Such corrections only affect lower-order terms $\E^1\B^1$ (Hall effect)
and $\E^2\B^0$ (nonlinear Hall effect), which are purely transverse;
therefore, the longitudinal resistivity remains unchanged.
In this manuscript, we mainly focus on the longitudinal effect.

\Eqr{sigma-ab}{sigma-mc'} are the needed ingredients to evaluate the
intrinsic contributions to the bulk eMChA tensor
$\gamma'_{abcd}$ in \eq{gamma-prime}. Inversion symmetry implies
$\epsilon(\kk)=\epsilon(-\kk)$, $\m(\kk)=\m(-\kk)$, and
$\OO(\kk)=\OO(-\kk)$; as a result, every term in
\eqs{sigma-Z}{sigma-mc'} vanishes in centrosymmetric crystals,
yielding $\gamma'_{abcd}=0$ as expected. Since
$\sigma_{ab}$ is linear in $\tau$ and $\sigma_{abcd}$ is quadratic,
$\gamma'_{abcd}$ becomes independent of
$\tau$ in the constant relaxation time approximation.

In the following, we first outline the computational methods,
and then report and discuss the numerical results of our
\textit{ab initio} calculations for $p$-Te.


\secl{Te}

\textit{Computational methods.}
The unit cell of trigonal Te contains three atoms disposed along a
spiral chain, with the chains arranged on a 2D hexagonal lattice.
The fully-relativistic electronic structure of the pristine (undoped)
right-handed crystal (space group P3$_1$21) is evaluated via
density-functional theory (DFT) in the pseudopotential framework, using the
HSE06 hybrid functional~\cite{paier-jcp06} implemented in the VASP
code package
\cite{kresse1999ultrasoft,kresse1996efficiency,PhysRevB.54.11169}.

In order to evaluate the needed $k$-space quantities accurately and
efficiently, we employ the Wannier interpolation
scheme~\cite{wang-prb06}.  The Wannier functions are constructed using
the Wannier90 code package~\cite{wannier90}, starting from
atom-centered $s$ and $p$ trial orbitals.  In order to avoid artifacts
due to possible numerical violations of crystal symmetries by the
Wannier model, the Wannier Hamiltonian and orbital-based matrix
elements are
symmetrized so as to satisfy the point-group symmetries.

We have implemented all terms in \eqs{sigma-Z}{sigma-mc'} in the
WannierBerri code package~\cite{wannierberri}.  Wannier interpolation
allows us to evaluate band properties and their momentum-space
derivatives directly, without finite difference
schemes 
~\cite{gradients}.
We perform the Brillouin-zone integrals on a symmetry-irreducible part
of a regular grid of $300\times300\times300$ points, using 30 adaptive
refinement iterations, and employing the tetrahedron 
method~\cite{tetrahedronmethod,tetrahedronmethod2}
to accurately describe the Fermi occupation factors \tenxl{$-f_0'$}.

\textit{Numerical results.}
The eMChA conductivity $\sigma_{abcd}$ is a nonmagnetic polar tensor
symmetric in $bc$, with the Jahn symbol
eV[V$^2$]V. Table~\ref{tab:Te-sigma-emcha-sym} shows the constraints
imposed on that tensor by the point group symmetry of trigonal
tellurium. 

\begin{table}[t]
\caption{Symmetry constraints on the components of the eMChA
    conductivity tensor $\sigma_{abcd}$ in crystal class
    32. The three largest calculated components in trigonal tellurium
    are highlighted in bold. The table was generated using the
$\rm MTENSOR$  routine~\cite{mTensor} provided by the
Bilbao Crystallographic Server~\cite{mTensor-link}.
}
\centering 
\makebox[\columnwidth]{\begin{tabular}{|c|c|cccccc|}
  \hline
  \multicolumn{2}{|c|}{32}& \multicolumn{6}{c|}{$bc=cb$}\\
  \cline{3-8}
    \multicolumn{2}{|c|}{$\sigma_{abcd}$} &$xx$&$yy$&$zz$&$yz$&$xz$&$xy$\\
  \hline
         &$xx$& $\sigma_{yyyy}$ & $\sigma_{yxxy}$& $\boldsymbol{\sigma_{yzzy}}$& $-\sigma_{yyzy}$& 0               & 0                 \\
         &$xy$& 0               & 0              & 0              & 0               & $-\sigma_{yyzy}$& $(\sigma_{yyyy}-\sigma_{yxxy})/2$ \\
         &$xz$& 0               & 0              & 0              & 0               & $\sigma_{yyzz}$ & $-\sigma_{yyyz}$  \\
         &$yx$& 0               & 0              & 0              & 0               & $-\sigma_{yyzy}$& $(\sigma_{yyyy}-\sigma_{yxxy})/2$ \\
 $ad$&$yy$& $\sigma_{yxxy}$ & $\sigma_{yyyy}$& $\boldsymbol{\sigma_{yzzy}}$& $\sigma_{yyzy}$ & 0               & 0                 \\
         &$yz$& $-\sigma_{zyyy}$& $\sigma_{yyyz}$& 0              & $\sigma_{yyzz}$ & 0               & 0                 \\
         &$zx$& 0               & 0              & 0              & 0               & $\sigma_{zyzy}$ & $-\sigma_{zyyy}$  \\
         &$zy$& $-\sigma_{zyyy}$& $\sigma_{zyyy}$& 0              & $\sigma_{zyzy}$ & 0               & 0                 \\
         &$zz$& $\sigma_{zyyz}$ & $\sigma_{zyyz}$& $\boldsymbol{\sigma_{zzzz}}$& 0               & 0               & 0                 \\
  \hline
\end{tabular}}
\tabl{Te-sigma-emcha-sym}
\end{table}

In $p$-Te the Fermi energy cuts the top of the upper valence
band shown in \fref{Te}(a), forming a small hole-like Fermi pocket
near the H point.  We find that when the Fermi level is placed near
the top of the upper valence band, the largest component of the
  eMChA tensor is
\beq
\gamma'_{zzzz}=-\rho^2_{zz}\sigma_{zzzz}
\eql{largest-gamma}
\eeq
and the second largest components are
\beq
\gamma'_{xzzx}=\gamma'_{yzzy}=-\rho_{yy}\rho_{zz}\sigma_{yzzy}
\eeq
(here $\rho_{xx}=\rho_{yy}$),
while other symmetry-allowed components are orders of
magnitude smaller (details in Supplementary Material \cite{SM}).  The
predominance of components driven by magnetic field 
along the trigonal $z$
axis suggests that the strong coupling to
$m_z$ is
crucial to the effect.
\begin{figure}[t]
\centering
\includegraphics[width=1\columnwidth]{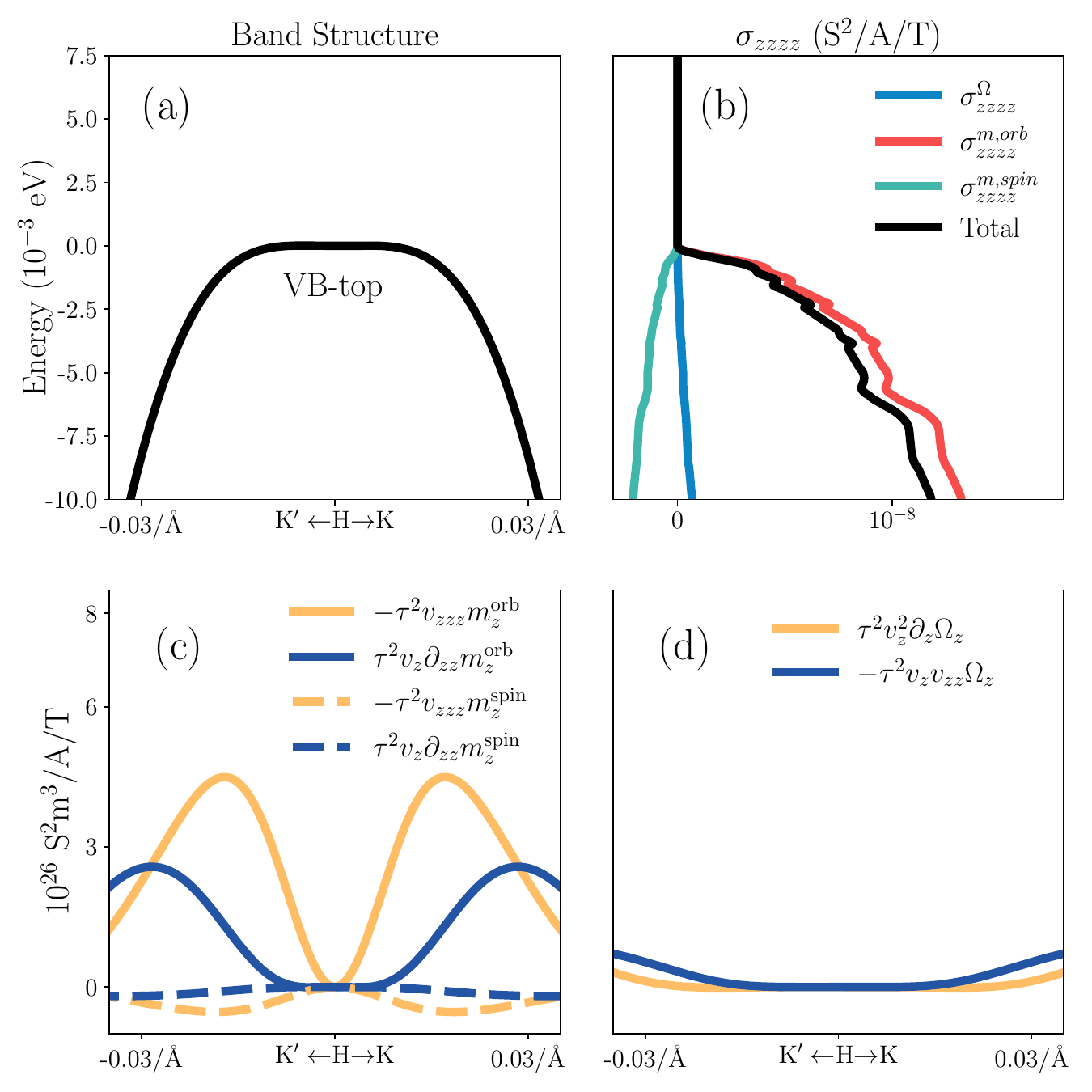}
\caption{(a) Upper valence band 
    of trigonal Te in a small segment of the K$^\prime$--H--K path
    around the H point. 
        (b) $zzzz$ component of the eMChA
    conductivity tensor as a function of chemical potential.
The total conductivity (black line) is broken down into
     contributions from the Berry curvature
  ($\Omega$), from the orbital Zeeman coupling
  (Z,orb), and from the spin Zeeman coupling
  (Z,spin). 
    (c)-(d) $k_z$
    dependent quantities in the conductivities $\sigma^{\rm Z}_{zzzz}$
  and $\sigma^\Omega_{zzzz}$ around H point alone principle axis
    $k_z$.  $\tau = 10^{-15}$s is used in (b),(c) and (d).
    }\figl{Te}
\end{figure}
As discussed earlier, the intrinsic eMChA conductivity has
three types of contributions:
orbital Zeeman, spin Zeeman, and Berry curvature.  For $p$-Te,
  the $zzzz$ component is strongly dominated by orbital Zeeman
contributions, as shown in \fref{Te}(b).  To understand this result,
consider the $zzzz$ components of
\eqs{sigma-Z}{sigma-mc'},
\begin{align}
\sigma^{\rm Z}_{zzzz}=&
    \tau^2 \int \dk \, (-v_{zzz}m_z + v_z \partial_{zz} m_z) (- f_0')\,,
\eql{z-z}\\
\sigma^{\rm \Omega}_{zzzz}=&  \tau^2\int \dk \, ( v^2_z\partial_z\Omega_z -
v_z v_{zz}\Omega_z  \nonumber\\
    &- 3 v_{zx}\Omega_x v_z - 3 v_{zy}\Omega_y v_z ) (-f_0') \eql{z-mc'}\,.
\end{align}
%
Due to the $-f'_0$ factors, only states located at the Fermi surface
contribute to the response at zero temperature.  Panels (c) and
  (d) of \fref{Te} show the evolution with $k_z$, near the H point,
of each term in the integrand of \eqs{z-z}{z-mc'}
(since $\Omega_x$ and $\Omega_y$ are negligible near  H,
we drop the terms in the second line of \eq{z-mc'}).

In \fref{Te}(c), the
contribution from each term in \eq{z-z} is further separated
  into orbital and spin parts; the orbital contributions are dominant,
  because $|m^{\rm orb}_z| \gg |m^{\rm spin}_z|$ in the upper valence
  band of Te~\cite{stepan18}. In each channel (spin or orbital), the
  first term in \eq{z-z} dominates over the second; this is due to the
  weak ``camel-back'' shape of the upper valence band near H, which
  suppresses the band gradient $v_z$ but not the higher derivative
  $v_{zzz}$. That also explains why the Berry-curvature contributions
  in \fref{Te}(d) are much smaller than the orbital-Zeeman ones in
  \fref{Te}(c), as both terms in the first line of \eq{z-mc'} contain
  $v_z$ factors.

\begin{figure}[t]
\centering
\includegraphics[width=1.0\columnwidth]{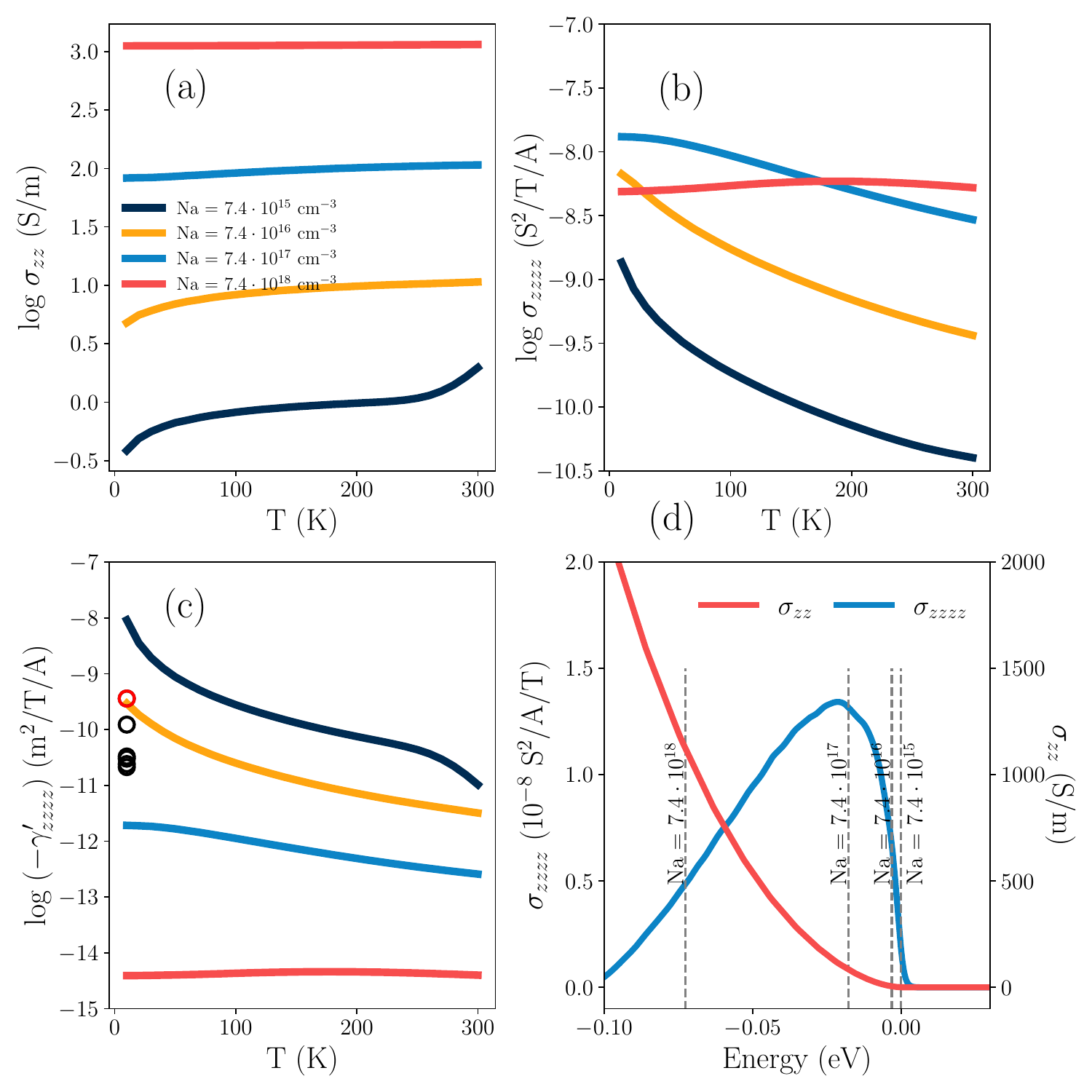}
\caption{Calculated temperature dependence, for different
    acceptor concentrations $N_{\rm a}$, of 
    (c) Ohmic conductivity $\sigma_{zz}$,
    (b) eMChA conductivity $\sigma_{zzzz}$, and
    (c) eMChA parameter $\gamma'_{zzzz}$. 
    $\tau = 10^{-15}$s is used in (a) and (b).
    The circles in panel~(c) denote the 
    values measured in six different samples in Ref.~\cite{calavalle2022gate};
    the red circle denotes the sample whose size was determined most
    precisely~\cite{calavalle2022gate}. (d) $\sigma_{zzzz}$ and
    $\sigma_{zz}$ as functions of chemical
    potential at 10K. The gray dashed lines indicate the chemical
  potentials corresponding to four different acceptor concentrations.
    }
\figl{T-dependent}
\end{figure}

The temperature-dependent $\sigma_{zz}$, 
$\sigma_{zzzz}$ and  $\gamma'_{zzzz}$ are shown in \fref{T-dependent}(a,b,c) respectively,
which are calculated using the temperature-dependent Fermi-Dirac
distribution function $f_0(T,\mu,\epsilon)$ at chemical potential $\mu$.
\fref{T-dependent}(d) displays $\sigma_{zzzz}$ and $\sigma_{zz}$ at four 
acceptor concentrations at 0K.
As the hole acceptor concentration increases in
$p$-Te, the Fermi level shifts downward, resulting in changes in
$\sigma_{zzzz}$ and $\sigma_{zz}$.  The specific Fermi level for
various acceptor concentrations is shown in \fref{T-dependent}(d).

When the Fermi level is close to the top of the valence band,
the acceptor
concentration is low, and the sensitivity of $\gamma'_{zzzz}$ to
temperature becomes significant compared to higher acceptor
concentrations.  In $\sigma_{zzzz}$, there is a high peak next to the
energy gap, leading to extreme changes due to temperature because of
the smearing of the peak.  The fact that $\sigma_{zzzz}$ and
$1/\sigma_{zz}$ have slopes in a similar trend that exacerbates 
the sensitivity of $\gamma'_{zzzz}$ to temperature when $T<50$ K.  When
$N_{\rm a} = 7.4\cdot10^{18}$cm$^{-3}$, 
the corresponding
 curve in \fref{T-dependent}(b) has an opposite slope
because the chemical potential
is on the other side of the peak of $\sigma_{zzzz}$ (see
\fref{T-dependent}(d)).
In the temperature dependent $\gamma'$ measurement in Ref.\cite{Hirobe22},
$\gamma' \approx 10^{-12}$m$^2$/T/A for gate voltage $V_g = 80$V at 20K.
From 20K to 160K, $\gamma'$ drops by one order of magnitude, see Fig.~4 and Fig.~2(a) therein. 
It matches our simulation when acceptor concentration is
between $7.4 \cdot 10^{16}$cm$^{-3}$ to $7.4 \cdot 10^{17}$cm$^{-3}$.

In the experimental measurement of $p$-Te nanowires in
Ref.\cite{calavalle2022gate}, eMChA was scanned over all possible
directions of the magnetic field, and it was confirmed that the
response is  largest when $\B$ is parallel to the applied current,
in agreement with our simulation.
The eMChA resistivity was
measured for six samples. However, to extract $\gamma'_{zzzz}$ the
exact dimensions
of the nanowires are needed, which were
determined only for one of them. Assuming that the size is the same
(at least in order of magnitude) for all nanowires, the
value of $\gamma'_{zzzz}$ for six nanowires at 10K may be estimated
within the range of
$[10^{-11},10^{-9}]~\text{m}^2/\text{T}/\text{A}$. The acceptor
concentration was estimated as $7.4\cdot10^{17} \text{cm}^{-3}$ using
Hall resistance $R_{\rm H}$,
\beq
N_{\rm a}  = \frac{B_c}{ewR_{\rm H}}= \frac{\sigma_{aa}\sigma_{bb}}{e\sigma^{H}_{abc}}  ~,
\eql{Na-Hall}
\eeq
where $w$ is the thickness
of a nanowire. It also can be evaluated from first 
principle by evaluating 
Ohmic and Hall conductivity $\sigma_{aa/bb}$ and $\sigma^{H}_{abc}$ 
within the Boltzman transport theory~\cite{hurd-book72}.
Strictly speaking, the first equality in \eq{Na-Hall} is only valid  for a parabolic band
dispersion. Even though the valence band of Te near the H point
is strongly non-parabolic, by means of accurate calculations we confirm that 
\eq{Na-Hall} is still valid close
to the top of the
valence band,
and the error only  increases slightly as the 
chemical potential moves away from the top of the valence band, 
see \fref{Na-gamma}(b). 
\begin{figure}[t]
\centering
\includegraphics[width=1.0\columnwidth]{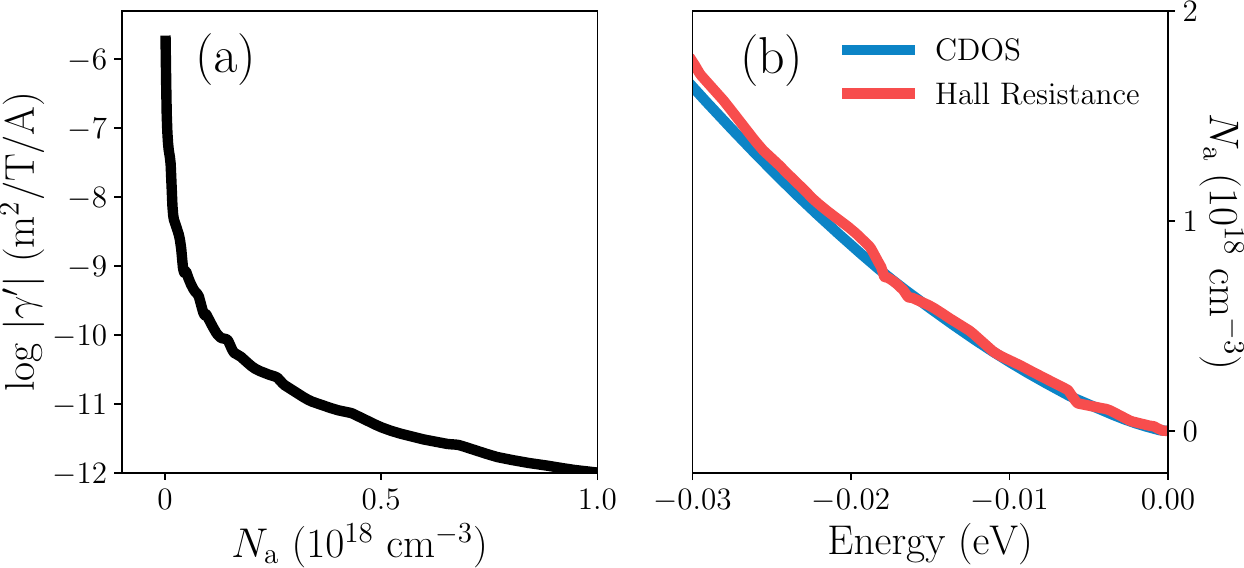}
    \caption{(a) Calculated $\gamma'_{zzzz}$ as a function of acceptor consentration. 
    (b) Comparison between acceptor concentration calculated using 
    precise cumulative density of states (CDOS) 
    and Hall resistance measurement.
    }
    \figl{Na-gamma}
\end{figure}
As shown in \fref{T-dependent}(a), our estimated value of
$\gamma'_{zzzz}$ is significantly lower for this acceptor
concentration. However, $\gamma'_{zzzz}$ is extremely
  sensitive to acceptor concentration.  A tiny change of $N_{\rm a}$
  changes orders of magnitude of $\gamma'_{zzzz}$, see
  \fref{Na-gamma}(a). Therefore, an excellent agreement with the experimental
value  is difficult to achieve.

In earlier measurements at room temperature \cite{Rikken19}, the
largest component of the $\gamma$ tensor was found to be in the range
of $[10^{-9},10^{-8}]~\text{m}^2/\text{T}/\text{A}$ with an acceptor
concentration of around $10^{16} \text{cm}^{-3}$. However, it was
observed that the effect is maximized when the magnetic field is
perpendicular to the current.  This contradicts the symmetry analysis
in Table \ref{tab:Te-sigma-emcha-sym}, and therefore cannot be
described as a bulk property of trigonal tellurium, regardless of the
microscopic mechanism implied.

  \textit{The sign of the effect.} Within constant relaxation time approximation we find that for the right-handed crystal, 
  the sign of the eMChA resistivity (the $\gamma'_{zzzz}$ parameter) is negative, or equivalently, the sign of the eMChA resistivity is positive.
  The opposite signs are expected for the left-handed crystal. This is consistent with the experimental results in Ref.\cite{calavalle2022gate}.
  However, after publication of the preprint of the present manuscript, another article appeared \cite{okumura2024chiralorbitaltexturenonlinear}, 
  that claims the opposite sign of the eMChA conductivity. The lack of information on the geometrical size of the samples in Ref.\cite{okumura2024chiralorbitaltexturenonlinear}
  does not allow us to make a direct comparison of the magnitude of the effect. It should be noted that the chirality of the crystal structure in the two experiments 
  was determined by different methods. In Ref.\cite{calavalle2022gate}, the chirality was determined by modern scanning transmission electron microscopy (STEM) technique\cite{dong2020atomic,ben2021chain}, 
  while in Ref.\cite{okumura2024chiralorbitaltexturenonlinear} it was determined by the more traditional method of the etch pits\cite{Koma1970etchpits}. 
  It is known that determination of chirality of trigonal tellurium is a non-trivial task, and the results of different methods are not always consistent (see appendix B of \cite{Furakawa-chirality} for a systematic analysis).
  On the theoretical part,  another article appeared recently~\cite{Golub2020semiclassical} showing that with a different choice of the collision integral
  (beyond the constant relaxation time approximation),
  the sign of the eMChA conductivity can be changed.

\textit{Summary.}
In this letter, we derived equations for quantifying eMChA
in real materials using the Boltzmann-transport formalism within the
constant relaxation time approximation.  We use $p$-Te as a research
platform and employ DFT and Wannier interpolation to make
theoretical predictions for eMChA conductivity $\sigma_{abcd}$ and
measurable tensor $\gamma'_{abcd}$.  We find that with low acceptor
concentrations, the orbital Zeeman coupling is the primary origin of
the eMChA response compared with spin Zeeman coupling and Berry
curvature contributions. 
We cannot boast of an excellent agreement with experimental data, 
but neither do we find consistency between different experiments.
Therefore, we hope that our results will be a good reference and motivation for further experimental and theoretical investigations.
The developed methodology is made
available via an open-source code in order to motivate further
first-principles
investigations of other materials.



\begin{acknowledgments}
\begin{sloppypar}
X.L. acknowledges financial support from the China Scholarship Council
and from the National Key R$\&$D Program of China (Grant
No. 2022YFA1403700). 
S.S.T. was funded by the European Union [H2020-MSCA-COFUND-2020-101034228-WOLFRAM2].
The work of I.S and S.S.T was funded by MCIN
MCIN/AEI/10.13039/501100011033 through Grant No. PID2021-129035NB-I00,
and by the European Union NextGenerationEU/PRTR-C17.I1, as well as by
the IKUR Strategy under the collaboration agreement between the
Donostia International Physics Center and the Materials Physics Center
on behalf of the Department of Education of the Basque Government.
The numerical calculations were supported by Center for Computational Science and Engineering of SUSTech.
%
%
We would like to thank G. L. J. A Rikken for useful discussions.
\end{sloppypar}
\end{acknowledgments}


\bibliography{bib}


\end{document}